\documentstyle[12pt,epsf]{article}
\newcommand{\mathrm}[1]{{\rm #1}}
\begin{document}
{\setlength{\parskip}{4pt plus 1pt}%
\flushbottom
\begin{titlepage}
\centerline{\large EUROPEAN LABORATORY FOR PARTICLE PHYSICS}
\bigskip
\begin{flushright}
CERN--PPE/97--090\\
18 July 1997\\
\end{flushright}
\vfill
\begin{center}
{\LARGE\bf\boldmath Updated measurement of the\\[2mm]
$\tau$ lepton lifetime}
\end{center}
\bigskip
\begin{center}
\renewcommand{\thefootnote}{\fnsymbol{footnote}}
\setcounter{footnote}{2}%
{\Large The ALEPH Collaboration}%
\footnote{See the following pages for the list of authors.}
\end{center}
\renewcommand{\thefootnote}{\arabic{footnote}}
\setcounter{footnote}{0}
\vfill
\begin{center}{\large Abstract}\end{center}
{\setlength\leftmargini{2.3em}
\begin{quote}
A new measurement of the mean lifetime of the $\tau$ lepton
is presented.
Three different analysis methods are applied to
a sample of 90000 $\tau$ pairs,
collected in 1993 and 1994 with the ALEPH detector at LEP.
The average of this measurement and
those previously published by ALEPH is
$\tau_\tau = 290.1 \pm 1.5 \pm 1.1 \,{\mathrm{fs}}$.
\end{quote}}
\vfill
\begin{center}
(To be published in Physics Letters B)
\end{center}
\bigskip\bigskip
\end{titlepage}
}%
\pagestyle{empty}
\newpage
\small
%
\newlength{\saveparskip}
\newlength{\savetextheight}
\newlength{\savetopmargin}
\newlength{\savetextwidth}
\newlength{\saveoddsidemargin}
\newlength{\savetopsep}
\newlength{\savehoffset}
\newlength{\savevoffset}
\setlength{\saveparskip}{\parskip}
\setlength{\savetextheight}{\textheight}
\setlength{\savetopmargin}{\topmargin}
\setlength{\savetextwidth}{\textwidth}
\setlength{\saveoddsidemargin}{\oddsidemargin}
\setlength{\savetopsep}{\topsep}
\setlength{\savehoffset}{\hoffset}
\setlength{\savevoffset}{\voffset}
%
%
\setlength{\parskip}{0.0cm}
\setlength{\textheight}{25.0cm}
\setlength{\topmargin}{-2.4cm}
\setlength{\textwidth}{16 cm}
\setlength{\oddsidemargin}{-0.0cm}
\setlength{\topsep}{1mm}
\setlength{\hoffset}{0mm}
\setlength{\voffset}{0mm}
\pretolerance=10000
\centerline{\large\bf The ALEPH Collaboration}
\footnotesize
\vspace{0.5cm}
{\raggedbottom
\begin{sloppypar}
\samepage\noindent
R.~Barate,
D.~Buskulic,
D.~Decamp,
P.~Ghez,
C.~Goy,
J.-P.~Lees,
A.~Lucotte,
M.-N.~Minard,
J.-Y.~Nief,
B.~Pietrzyk
\nopagebreak
\begin{center}
\parbox{15.5cm}{\sl\samepage
Laboratoire de Physique des Particules (LAPP), IN$^{2}$P$^{3}$-CNRS,
74019 Annecy-le-Vieux Cedex, France}
\end{center}\end{sloppypar}
\vspace{2mm}
\begin{sloppypar}
\noindent
M.P.~Casado,
M.~Chmeissani,
P.~Comas,
J.M.~Crespo,
M.~Delfino,
E.~Fernandez,
M.~Fernandez-Bosman,
Ll.~Garrido,$^{15}$
A.~Juste,
M.~Martinez,
G.~Merino,
R.~Miquel,
Ll.M.~Mir,
C.~Padilla,
I.C.~Park,
A.~Pascual,
J.A.~Perlas,
I.~Riu,
F.~Sanchez,
F.~Teubert
\nopagebreak
\begin{center}
\parbox{15.5cm}{\sl\samepage
Institut de F\'{i}sica d'Altes Energies, Universitat Aut\`{o}noma
de Barcelona, 08193 Bellaterra (Barcelona), Spain$^{7}$}
\end{center}\end{sloppypar}
\vspace{2mm}
\begin{sloppypar}
\noindent
A.~Colaleo,
D.~Creanza,
M.~de~Palma,
G.~Gelao,
G.~Iaselli,
G.~Maggi,
M.~Maggi,
N.~Marinelli,
S.~Nuzzo,
A.~Ranieri,
G.~Raso,
F.~Ruggieri,
G.~Selvaggi,
L.~Silvestris,
P.~Tempesta,
A.~Tricomi,$^{3}$
G.~Zito
\nopagebreak
\begin{center}
\parbox{15.5cm}{\sl\samepage
Dipartimento di Fisica, INFN Sezione di Bari, 70126
Bari, Italy}
\end{center}\end{sloppypar}
\vspace{2mm}
\begin{sloppypar}
\noindent
X.~Huang,
J.~Lin,
Q. Ouyang,
T.~Wang,
Y.~Xie,
R.~Xu,
S.~Xue,
J.~Zhang,
L.~Zhang,
W.~Zhao
\nopagebreak
\begin{center}
\parbox{15.5cm}{\sl\samepage
Institute of High-Energy Physics, Academia Sinica, Beijing, The People's
Republic of China$^{8}$}
\end{center}\end{sloppypar}
\vspace{2mm}
\begin{sloppypar}
\noindent
D.~Abbaneo,
R.~Alemany,
U.~Becker,
A.O.~Bazarko,$^{20}$
P.~Bright-Thomas,
M.~Cattaneo,
F.~Cerutti,
G.~Dissertori,
H.~Drevermann,
R.W.~Forty,
M.~Frank,
R.~Hagelberg,
J.B.~Hansen,
J.~Harvey,
P.~Janot,
B.~Jost,
E.~Kneringer,
J.~Knobloch,
I.~Lehraus,
P.~Mato,
A.~Minten,
L.~Moneta,
A.~Pacheco,
J.-F.~Pusztaszeri,$^{23}$
F.~Ranjard,
G.~Rizzo,
L.~Rolandi,
D.~Rousseau,
D.~Schlatter,
M.~Schmitt,
O.~Schneider,
W.~Tejessy,
I.R.~Tomalin,
H.~Wachsmuth,
A.~Wagner$^{24}$
\nopagebreak
\begin{center}
\parbox{15.5cm}{\sl\samepage
European Laboratory for Particle Physics (CERN), 1211 Geneva 23,
Switzerland}
\end{center}\end{sloppypar}
\vspace{2mm}
\begin{sloppypar}
\noindent
Z.~Ajaltouni,
A.~Barr\`{e}s,
C.~Boyer,
A.~Falvard,
C.~Ferdi,
P.~Gay,
C~.~Guicheney,
P.~Henrard,
J.~Jousset,
B.~Michel,
S.~Monteil,
J-C.~Montret,
D.~Pallin,
P.~Perret,
F.~Podlyski,
J.~Proriol,
P.~Rosnet,
J.-M.~Rossignol
\nopagebreak
\begin{center}
\parbox{15.5cm}{\sl\samepage
Laboratoire de Physique Corpusculaire, Universit\'e Blaise Pascal,
IN$^{2}$P$^{3}$-CNRS, Clermont-Ferrand, 63177 Aubi\`{e}re, France}
\end{center}\end{sloppypar}
\vspace{2mm}
\begin{sloppypar}
\noindent
T.~Fearnley,
J.D.~Hansen,
J.R.~Hansen,
P.H.~Hansen,
B.S.~Nilsson,
B.~Rensch,
A.~W\"a\"an\"anen
\begin{center}
\parbox{15.5cm}{\sl\samepage
Niels Bohr Institute, 2100 Copenhagen, Denmark$^{9}$}
\end{center}\end{sloppypar}
\vspace{2mm}
\begin{sloppypar}
\noindent
G.~Daskalakis,
A.~Kyriakis,
C.~Markou,
E.~Simopoulou,
I.~Siotis,
A.~Vayaki
\nopagebreak
\begin{center}
\parbox{15.5cm}{\sl\samepage
Nuclear Research Center Demokritos (NRCD), Athens, Greece}
\end{center}\end{sloppypar}
\vspace{2mm}
\begin{sloppypar}
\noindent
A.~Blondel,
G.~Bonneaud,
J.C.~Brient,
P.~Bourdon,
A.~Roug\'{e},
M.~Rumpf,
A.~Valassi,$^{6}$
M.~Verderi,
H.~Videau
\nopagebreak
\begin{center}
\parbox{15.5cm}{\sl\samepage
Laboratoire de Physique Nucl\'eaire et des Hautes Energies, Ecole
Polytechnique, IN$^{2}$P$^{3}$-CNRS, 91128 Palaiseau Cedex, France}
\end{center}\end{sloppypar}
\vspace{2mm}
\begin{sloppypar}
\noindent
D.J.~Candlin,
M.I.~Parsons
\nopagebreak
\begin{center}
\parbox{15.5cm}{\sl\samepage
Department of Physics, University of Edinburgh, Edinburgh EH9 3JZ,
 United Kingdom$^{10}$}
\end{center}\end{sloppypar}
\vspace{2mm}
\begin{sloppypar}
\noindent
E.~Focardi,
G.~Parrini,
K.~Zachariadou
\nopagebreak
\begin{center}
\parbox{15.5cm}{\sl\samepage
Dipartimento di Fisica, Universit\`a di Firenze, INFN Sezione di Firenze,
50125 Firenze, Italy}
\end{center}\end{sloppypar}
\vspace{2mm}
\begin{sloppypar}
\noindent
M.~Corden,
C.~Georgiopoulos,
D.E.~Jaffe
\nopagebreak
\begin{center}
\parbox{15.5cm}{\sl\samepage
Supercomputer Computations Research Institute,
Florida State University,
Tallahassee, FL 32306-4052, USA $^{13,14}$}
\end{center}\end{sloppypar}
\vspace{2mm}
\begin{sloppypar}
\noindent
A.~Antonelli,
G.~Bencivenni,
G.~Bologna,$^{4}$
F.~Bossi,
P.~Campana,
G.~Capon,
D.~Casper,
V.~Chiarella,
G.~Felici,
P.~Laurelli,
G.~Mannocchi,$^{5}$
F.~Murtas,
G.P.~Murtas,
L.~Passalacqua,
M.~Pepe-Altarelli
\nopagebreak
\begin{center}
\parbox{15.5cm}{\sl\samepage
Laboratori Nazionali dell'INFN (LNF-INFN), 00044 Frascati, Italy}
\end{center}\end{sloppypar}
\vspace{2mm}
\begin{sloppypar}
\noindent
L.~Curtis,
S.J.~Dorris,
A.W.~Halley,
I.G.~Knowles,
J.G.~Lynch,
V.~O'Shea,
C.~Raine,
J.M.~Scarr,
K.~Smith,
P.~Teixeira-Dias,
A.S.~Thompson,
E.~Thomson,
F.~Thomson,
R.M.~Turnbull
\nopagebreak
\begin{center}
\parbox{15.5cm}{\sl\samepage
Department of Physics and Astronomy, University of Glasgow, Glasgow G12
8QQ,United Kingdom$^{10}$}
\end{center}\end{sloppypar}
\vspace{2mm}
\begin{sloppypar}
\noindent
O.~Buchm\"uller,
S.~Dhamotharan,
C.~Geweniger,
G.~Graefe,
P.~Hanke,
G.~Hansper,
V.~Hepp,
E.E.~Kluge,
A.~Putzer,
J.~Sommer,
K.~Tittel,
S.~Werner,
M.~Wunsch
\nopagebreak
\begin{center}
\parbox{15.5cm}{\sl\samepage
Institut f\"ur Hochenergiephysik, Universit\"at Heidelberg, 69120
Heidelberg, Fed.\ Rep.\ of Germany$^{16}$}
\end{center}\end{sloppypar}
\vspace{2mm}
\begin{sloppypar}
\noindent
R.~Beuselinck,
D.M.~Binnie,
W.~Cameron,
P.J.~Dornan,
M.~Girone,
S.~Goodsir,
E.B.~Martin,
A.~Moutoussi,
J.~Nash,
J.K.~Sedgbeer,
P.~Spagnolo,
A.M.~Stacey,
M.D.~Williams
\nopagebreak
\begin{center}
\parbox{15.5cm}{\sl\samepage
Department of Physics, Imperial College, London SW7 2BZ,
United Kingdom$^{10}$}
\end{center}\end{sloppypar}
\vspace{2mm}
\begin{sloppypar}
\noindent
V.M.~Ghete,
P.~Girtler,
D.~Kuhn,
G.~Rudolph
\nopagebreak
\begin{center}
\parbox{15.5cm}{\sl\samepage
Institut f\"ur Experimentalphysik, Universit\"at Innsbruck, 6020
Innsbruck, Austria$^{18}$}
\end{center}\end{sloppypar}
\vspace{2mm}
\begin{sloppypar}
\noindent
A.P.~Betteridge,
C.K.~Bowdery,
P.~Colrain,
G.~Crawford,
A.J.~Finch,
F.~Foster,
G.~Hughes,
R.W.L.~Jones,
T.~Sloan,
M.I.~Williams
\nopagebreak
\begin{center}
\parbox{15.5cm}{\sl\samepage
Department of Physics, University of Lancaster, Lancaster LA1 4YB,
United Kingdom$^{10}$}
\end{center}\end{sloppypar}
\vspace{2mm}
\begin{sloppypar}
\noindent
A.~Galla,
I.~Giehl,
A.M.~Greene,
C.~Hoffmann,
K.~Jakobs,
K.~Kleinknecht,
G.~Quast,
B.~Renk,
E.~Rohne,
H.-G.~Sander,
P.~van~Gemmeren,
C.~Zeitnitz
\nopagebreak
\begin{center}
\parbox{15.5cm}{\sl\samepage
Institut f\"ur Physik, Universit\"at Mainz, 55099 Mainz, Fed.\ Rep.\
of Germany$^{16}$}
\end{center}\end{sloppypar}
\vspace{2mm}
\begin{sloppypar}
\noindent
J.J.~Aubert,
C.~Benchouk,
A.~Bonissent,
G.~Bujosa,
J.~Carr,
P.~Coyle,
C.~Diaconu,
F.~Etienne,
N.~Konstantinidis,
O.~Leroy,
F.~Motsch,
P.~Payre,
M.~Talby,
A.~Sadouki,
M.~Thulasidas,
K.~Trabelsi
\nopagebreak
\begin{center}
\parbox{15.5cm}{\sl\samepage
Centre de Physique des Particules, Facult\'e des Sciences de Luminy,
IN$^{2}$P$^{3}$-CNRS, 13288 Marseille, France}
\end{center}\end{sloppypar}
\vspace{2mm}
\begin{sloppypar}
\noindent
M.~Aleppo,
M.~Antonelli,
F.~Ragusa$^{2}$
\nopagebreak
\begin{center}
\parbox{15.5cm}{\sl\samepage
Dipartimento di Fisica, Universit\`a di Milano e INFN Sezione di Milano,
20133 Milano, Italy}
\end{center}\end{sloppypar}
\vspace{2mm}
\begin{sloppypar}
\noindent
R.~Berlich,
W.~Blum,
V.~B\"uscher,
H.~Dietl,
G.~Ganis,
C.~Gotzhein,
H.~Kroha,
G.~L\"utjens,
G.~Lutz,
W.~M\"anner,
H.-G.~Moser,
R.~Richter,
A.~Rosado-Schlosser,
S.~Schael,
R.~Settles,
H.~Seywerd,
R.~St.~Denis,
H.~Stenzel,
W.~Wiedenmann,
G.~Wolf
\nopagebreak
\begin{center}
\parbox{15.5cm}{\sl\samepage
Max-Planck-Institut f\"ur Physik, Werner-Heisenberg-Institut,
80805 M\"unchen, Fed.\ Rep.\ of Germany\footnotemark[16]}
\end{center}\end{sloppypar}
\vspace{2mm}
\begin{sloppypar}
\noindent
J.~Boucrot,
O.~Callot,$^{2}$
S.~Chen,
Y.~Choi,$^{21}$
A.~Cordier,
M.~Davier,
L.~Duflot,
J.-F.~Grivaz,
Ph.~Heusse,
A.~H\"ocker,
A.~Jacholkowska,
M.~Jacquet,
D.W.~Kim,$^{12}$
F.~Le~Diberder,
J.~Lefran\c{c}ois,
A.-M.~Lutz,
I.~Nikolic,
M.-H.~Schune,
S.~Simion,
E.~Tournefier,
J.-J.~Veillet,
I.~Videau,
D.~Zerwas
\nopagebreak
\begin{center}
\parbox{15.5cm}{\sl\samepage
Laboratoire de l'Acc\'el\'erateur Lin\'eaire, Universit\'e de Paris-Sud,
IN$^{2}$P$^{3}$-CNRS, 91405 Orsay Cedex, France}
\end{center}\end{sloppypar}
\vspace{2mm}
\begin{sloppypar}
\noindent
\samepage
P.~Azzurri,
G.~Bagliesi,
G.~Batignani,
S.~Bettarini,
C.~Bozzi,
G.~Calderini,
M.~Carpinelli,
M.A.~Ciocci,
V.~Ciulli,
R.~Dell'Orso,
R.~Fantechi,
I.~Ferrante,
L.~Fo\`{a},$^{1}$
F.~Forti,
A.~Giassi,
M.A.~Giorgi,
A.~Gregorio,
F.~Ligabue,
A.~Lusiani,
P.S.~Marrocchesi,
A.~Messineo,
F.~Palla,
G.~Sanguinetti,
A.~Sciab\`a,
J.~Steinberger,
R.~Tenchini,
G.~Tonelli,$^{19}$
C.~Vannini,
A.~Venturi,
P.G.~Verdini
\samepage
\begin{center}
\parbox{15.5cm}{\sl\samepage
Dipartimento di Fisica dell'Universit\`a, INFN Sezione di Pisa,
e Scuola Normale Superiore, 56010 Pisa, Italy}
\end{center}\end{sloppypar}
\vspace{2mm}
\begin{sloppypar}
\noindent
G.A.~Blair,
L.M.~Bryant,
J.T.~Chambers,
Y.~Gao,
M.G.~Green,
T.~Medcalf,
P.~Perrodo,
J.A.~Strong,
J.H.~von~Wimmersperg-Toeller
\nopagebreak
\begin{center}
\parbox{15.5cm}{\sl\samepage
Department of Physics, Royal Holloway \& Bedford New College,
University of London, Surrey TW20 OEX, United Kingdom$^{10}$}
\end{center}\end{sloppypar}
\vspace{2mm}
\begin{sloppypar}
\noindent
D.R.~Botterill,
R.W.~Clifft,
T.R.~Edgecock,
S.~Haywood,
P.R.~Norton,
J.C.~Thompson,
A.E.~Wright
\nopagebreak
\begin{center}
\parbox{15.5cm}{\sl\samepage
Particle Physics Dept., Rutherford Appleton Laboratory,
Chilton, Didcot, Oxon OX11 OQX, United Kingdom$^{10}$}
\end{center}\end{sloppypar}
\vspace{2mm}
\begin{sloppypar}
\noindent
B.~Bloch-Devaux,
P.~Colas,
S.~Emery,
W.~Kozanecki,
E.~Lan\c{c}on,
M.C.~Lemaire,
E.~Locci,
P.~Perez,
J.~Rander,
J.-F.~Renardy,
A.~Roussarie,
J.-P.~Schuller,
J.~Schwindling,
A.~Trabelsi,
B.~Vallage
\nopagebreak
\begin{center}
\parbox{15.5cm}{\sl\samepage
CEA, DAPNIA/Service de Physique des Particules,
CE-Saclay, 91191 Gif-sur-Yvette Cedex, France$^{17}$}
\end{center}\end{sloppypar}
\pagebreak
\vspace{2mm}
\begin{sloppypar}
\noindent
S.N.~Black,
J.H.~Dann,
R.P.~Johnson,
H.Y.~Kim,
A.M.~Litke,
M.A. McNeil,
G.~Taylor
\nopagebreak
\begin{center}
\parbox{15.5cm}{\sl\samepage
Institute for Particle Physics, University of California at
Santa Cruz, Santa Cruz, CA 95064, USA$^{22}$}
\end{center}\end{sloppypar}
\vspace{2mm}
\begin{sloppypar}
\noindent
C.N.~Booth,
R.~Boswell,
C.A.J.~Brew,
S.~Cartwright,
F.~Combley,
M.S.~Kelly,
M.~Lehto,
W.M.~Newton,
J.~Reeve,
L.F.~Thompson
\nopagebreak
\begin{center}
\parbox{15.5cm}{\sl\samepage
Department of Physics, University of Sheffield, Sheffield S3 7RH,
United Kingdom$^{10}$}
\end{center}\end{sloppypar}
\vspace{2mm}
\begin{sloppypar}
\noindent
A.~B\"ohrer,
S.~Brandt,
G.~Cowan,
C.~Grupen,
G.~Lutters,
P.~Saraiva,
L.~Smolik,
F.~Stephan
\nopagebreak
\begin{center}
\parbox{15.5cm}{\sl\samepage
Fachbereich Physik, Universit\"at Siegen, 57068 Siegen,
 Fed.\ Rep.\ of Germany$^{16}$}
\end{center}\end{sloppypar}
\vspace{2mm}
\begin{sloppypar}
\noindent
M.~Apollonio,
L.~Bosisio,
R.~Della~Marina,
G.~Giannini,
B.~Gobbo,
G.~Musolino
\nopagebreak
\begin{center}
\parbox{15.5cm}{\sl\samepage
Dipartimento di Fisica, Universit\`a di Trieste e INFN Sezione di Trieste,
34127 Trieste, Italy}
\end{center}\end{sloppypar}
\vspace{2mm}
\begin{sloppypar}
\noindent
J.~Putz,
J.~Rothberg,
S.~Wasserbaech
\nopagebreak
\begin{center}
\parbox{15.5cm}{\sl\samepage
Experimental Elementary Particle Physics, University of Washington, WA 98195
Seattle, U.S.A.}
\end{center}\end{sloppypar}
\vspace{2mm}
\begin{sloppypar}
\noindent
S.R.~Armstrong,
E.~Charles,
P.~Elmer,
D.P.S.~Ferguson,
S.~Gonz\'{a}lez,
T.C.~Greening,
O.J.~Hayes,
H.~Hu,
S.~Jin,
P.A.~McNamara III,
J.M.~Nachtman,
J.~Nielsen,
W.~Orejudos,
Y.B.~Pan,
Y.~Saadi,
I.J.~Scott,
J.~Walsh,
Sau~Lan~Wu,
X.~Wu,
J.M.~Yamartino,
G.~Zobernig
\nopagebreak
\begin{center}
\parbox{15.5cm}{\sl\samepage
Department of Physics, University of Wisconsin, Madison, WI 53706,
USA$^{11}$}
\end{center}\end{sloppypar}
}
\footnotetext[1]{Now at CERN, 1211 Geneva 23,
Switzerland.}
\footnotetext[2]{Also at CERN, 1211 Geneva 23, Switzerland.}
\footnotetext[3]{Also at Dipartimento di Fisica, INFN, Sezione di Catania, Catania, Italy.}
\footnotetext[4]{Also Istituto di Fisica Generale, Universit\`{a} di
Torino, Torino, Italy.}
\footnotetext[5]{Also Istituto di Cosmo-Geofisica del C.N.R., Torino,
Italy.}
\footnotetext[6]{Supported by the Commission of the European Communities,
contract ERBCHBICT941234.}
\footnotetext[7]{Supported by CICYT, Spain.}
\footnotetext[8]{Supported by the National Science Foundation of China.}
\footnotetext[9]{Supported by the Danish Natural Science Research Council.}
\footnotetext[10]{Supported by the UK Particle Physics and Astronomy Research
Council.}
\footnotetext[11]{Supported by the US Department of Energy, grant
DE-FG0295-ER40896.}
\footnotetext[12]{Permanent address: Kangnung National University, Kangnung, 
Korea.}
\footnotetext[13]{Supported by the US Department of Energy, contract
DE-FG05-92ER40742.}
\footnotetext[14]{Supported by the US Department of Energy, contract
DE-FC05-85ER250000.}
\footnotetext[15]{Permanent address: Universitat de Barcelona, 08208 Barcelona,
Spain.}
\footnotetext[16]{Supported by the Bundesministerium f\"ur Bildung,
Wissenschaft, Forschung und Technologie, Fed. Rep. of Germany.}
\footnotetext[17]{Supported by the Direction des Sciences de la
Mati\`ere, C.E.A.}
\footnotetext[18]{Supported by Fonds zur F\"orderung der wissenschaftlichen
Forschung, Austria.}
\footnotetext[19]{Also at Istituto di Matematica e Fisica,
Universit\`a di Sassari, Sassari, Italy.}
\footnotetext[20]{Now at Princeton University, Princeton, NJ 08544, U.S.A.}
\footnotetext[21]{Permanent address: Sung Kyun Kwan University, Suwon, Korea.}
\footnotetext[22]{Supported by the US Department of Energy,
grant DE-FG03-92ER40689.}
\footnotetext[23]{Now at School of Operations Research and Industrial
Engineering, Cornell University, Ithaca, NY 14853-3801, U.S.A.}
\footnotetext[24]{Now at Schweizerischer Bankverein, Basel, Switzerland.}
%
%
\setlength{\parskip}{\saveparskip}
\setlength{\textheight}{\savetextheight}
\setlength{\topmargin}{\savetopmargin}
\setlength{\textwidth}{\savetextwidth}
\setlength{\oddsidemargin}{\saveoddsidemargin}
\setlength{\topsep}{\savetopsep}
\setlength{\hoffset}{\savehoffset}
\setlength{\voffset}{\savevoffset}
\normalsize
\newpage
\pagestyle{plain}
\setcounter{page}{1}
\setlength{\parskip}{4pt plus 1pt}%
\raggedbottom%
\section{Introduction}
\label{s:intro}%
In the Standard Model,
the weak charged-current coupling strength is assumed to 
be the same for all fermion doublets.
This hypothesis may be tested in the lepton sector
by comparing the rates of certain decays.
The most precise universality tests
involving the $\tau$-$\nu_\tau$ doublet
are presently obtained from measurements of
the decays $\tau^-\rightarrow e^-\nu\bar{\nu}$, 
$\tau^-\rightarrow\mu^-\nu\bar{\nu}$, and 
$\mu^-\rightarrow e^-\nu\bar{\nu}$,
the sensitivity being limited by the experimental
uncertainties on the $\tau$ lifetime and branching fractions.

Presented in this paper is
an updated measurement of the $\tau$ lifetime,
based on the momentum-dependent impact parameter sum (MIPS) method,
the impact parameter difference (IPD) method,
and the decay length (DL) method.
The MIPS and IPD analyses are applied to the 
$e^+e^-\rightarrow\tau^+\tau^-$ events
in which each $\tau$ decays to a final state containing
a single charged track (``1-1 topology'').
The MIPS measurement has a small statistical uncertainty
because the impact parameter smearing related to the size
of the luminous region is nearly canceled in the
sum of the impact parameters of the two daughter tracks.
The result is, however, sensitive to the
assumed impact parameter resolution.
On the other hand, the IPD measurement is subject to a
statistical error from the size of the luminous region
but is insensitive to the assumed resolution.
The DL method yields a precise lifetime measurement
from $\tau$'s decaying into three-prong final states.
The dominant source of uncertainty is 
the statistical uncertainty
related to the natural width of the exponential lifetime
distribution;
the size of the luminous region
yields a negligible contribution to the lifetime uncertainty.
Brief descriptions of
the MIPS, IPD, and DL methods are given below;
more details are available in~[\hbox{1}--\hbox{3}].
The ALEPH measurement of the $\tau$ lifetime is
further supplemented by an analysis~\cite{alephthreedip}
based on the three-dimensional impact parameter sum (3DIP) method.
The 3DIP method relies on kinematic constraints to
reduce the $\tau$ direction uncertainty in events
containing two hadronic $\tau$ decays.

The data were collected at the LEP $e^+e^-$ collider,
at centre of mass energies around the Z resonance.
A $\tau$ mass of
$m_\tau = 1777.00^{+0.30}_{-0.27} \,{\mathrm{MeV}}/c^2$~\cite{rpp}
is assumed throughout this paper.

\section{Apparatus and data sample}
\label{s:apparatus}%
The ALEPH detector is described in detail in~[\hbox{6},\hbox{7}].
The tracking system consists of
a high-resolution silicon strip vertex detector (VDET),
a cylindrical drift chamber (the inner tracking chamber or ITC),
and a large time projection chamber (TPC).
The VDET comprises two layers
of double-sided silicon strip detectors
at average radii of $6.3$ and $10.8\,{\mathrm{cm}}$.
The spatial resolution for $r$-$\phi$ coordinates is 
$12 \,\mu{\mathrm{m}}$
and varies between 12 and $22\,\mu{\mathrm{m}}$ for $z$ coordinates,
depending on track polar angle.
The angular coverage is
$\left|\cos\theta\right| < 0.85$ for the inner layer and
$\left|\cos\theta\right| < 0.69$ for the outer layer.
The ITC has eight coaxial wire layers
at radii from 16 to $26\,{\mathrm{cm}}$.
The TPC provides up to 21 three-dimensional coordinates per track
at radii between 40 and $171\,{\mathrm{cm}}$.
The tracking detectors are contained within
a superconducting solenoid,
which produces an axial magnetic field of $1.5\,{\mathrm{T}}$.
Charged tracks measured in the VDET-ITC-TPC system
are reconstructed with a momentum resolution of
$\Delta p/p = 
6\times 10^{-4} \, p_{t} \oplus 0.005$ 
($p_{t}$ in ${\mathrm{GeV}}/c$).
An impact parameter resolution of $28\,\mu{\mathrm{m}}$
in the $r$-$\phi$ plane is achieved for muons from 
${\mathrm{Z}}\rightarrow\mu^+\mu^-$ having at least
one VDET $r$-$\phi$ hit.

Surrounding the TPC is an electromagnetic calorimeter (ECAL),
a lead/wire-chamber sandwich operated in proportional mode.
The calorimeter is read out via projective towers
subtending typically $0.9^{\circ} \times 0.9^{\circ}$ in solid angle
which sum the deposited energy in three sections in depth.
Beyond the ECAL lies the solenoid,
followed by a hadron calorimeter (HCAL),
which uses the iron return yoke as absorber
and has an average depth of $1.50\,{\mathrm{m}}$.
Hadronic showers are sampled by 23 planes of streamer tubes,
providing a digital hit pattern and
inducing an analog signal on pads arranged in projective towers.
The HCAL is used in combination with two layers of muon chambers
outside the magnet for $\mu$ identification.

The analysis is based on data samples collected in 1993 and 1994,
corresponding to an estimated 31900 and 82700
produced $\tau$ pairs, respectively.
The combined sample of $e^+e^-\rightarrow\tau^+\tau^-$
events consists of
3.5\% collected at $\sqrt{s} = 89.4 \,{\mathrm{GeV}}$,
91.4\% at $91.2 \,{\mathrm{GeV}}$ (Z ``peak''), and
5.1\% at $93.0 \,{\mathrm{GeV}}$.
All of the off-peak data were obtained in 1993.
The data from the two years are analyzed separately.
No inconsistencies between the samples are observed.
The results presented herein refer to the combination
of the two data sets.

Tau-pair candidate events are selected by means of
an improved version of the algorithm described in~\cite{tsltzeroun}.
The modifications make the selection less sensitive to
low-energy clusters in the calorimeters.
The overall efficiency of this algorithm is $78\%$,
with an expected background contamination of $1.3\%$
at the Z peak.
A total of 90408 candidate $\tau^+\tau^-$ events are selected.
Further selection criteria
are then imposed to isolate well-reconstructed
one-prong and three-prong $\tau$ decays
for the different lifetime analyses.

Monte Carlo $\tau^+\tau^-$ events~\cite{koralz}
are used to study biases in the analysis methods.
An independent sample is generated for each year of data taking
to simulate the applicable
detector conditions, beam sizes, and
centre of mass energies.
The input $\tau$ lifetime is $296\,{\mathrm{fs}}$.
Background events from
$e^+e^-\rightarrow e^+e^-$~\cite{ubab},
$e^+e^-\rightarrow\mu^+\mu^-$~\cite{koralz},
$e^+e^-\rightarrow\mbox{\sl q}\bar{\mbox{\sl q}}$~\cite{hvfl}, and
two-photon interactions~\cite{phot-ggmj}
are also simulated.

In the following, $d$ denotes
the impact parameter of a reconstructed daughter track,
measured in the $r$-$\phi$ plane
with respect to the nominal interaction point.
The sign of $d$ is defined to be that of the $z$ component
of the particle's angular momentum about this point.

Reconstructed track impact parameters are corrected for
systematic offsets in both $d$ and $z$
due to detector alignment and drift field parametrization errors.
The offsets are measured as a function of $\theta$ and $\phi$
from ${\mathrm{Z}}\rightarrow\mbox{\sl q}\bar{\mbox{\sl q}}$ events.
The corrections have an rms of $14\,\mu{\mathrm{m}}$ in $d$ and
$20\,\mu{\mathrm{m}}$ in $z$.
Similar results are obtained from 
${\mathrm{Z}}\rightarrow\mu^+\mu^-$ events,
confirming that the offsets do not depend appreciably
on track momentum.
These corrections have a small effect and are taken into account
in the evaluation of the systematic uncertainty in each analysis.

The beam axis position in the $xy$ plane
is determined from selected reconstructed
charged tracks in Z decay events (excluding $\tau$ pairs),
averaged over blocks of roughly 75 consecutive events.
The typical uncertainties are $20\,\mu{\mathrm{m}}$ 
in $x$ (horizontally)
and $10\,\mu{\mathrm{m}}$ in $y$ (vertically).
The size of the luminous region along $x$
(typically $159\,\mu{\mathrm{m}}$ rms in 1993
and $125\,\mu{\mathrm{m}}$ in 1994,
due to different LEP optics)
is measured from the fitted primary vertices of selected 
${\mathrm{Z}}\rightarrow\mbox{\sl q}\bar{\mbox{\sl q}}$ events,
over blocks of about 270 events.
The size in the $y$ direction is taken to be
$5\,\mu{\mathrm{m}}$ rms.
The measured $z$ dimensions,
$7.4\,{\mathrm{mm}}$ in 1993 and $7.0\,{\mathrm{mm}}$ in 1994,
are used in the DL analysis.

\section{Momentum-dependent impact parameter sum\newline analysis}
\label{s:mips}%
The MIPS method~\cite{alephiii}
is applied to $\tau^+\tau^-$ events of 1-1 topology.
The mean $\tau$ lifetime is measured by means of
a maximum likelihood fit to the distribution of the
impact parameter sum $\delta = d_+ +d_-$, 
where $d_+$ and $d_-$ denote the impact parameters
of the two charged daughter tracks.
The mean lifetime determines the overall scale of the
true $\delta$ distribution;
typical $\delta$ values are on the order of
$c\tau_\tau \simeq 88\,\mu{\mathrm{m}}$.
The likelihood function is constructed to take into account,
on an event-by-event basis,
the dependence of the $\delta$ distribution on
the momenta $p_\pm$
of the daughters.
A large sample of Monte Carlo events is used to
parametrize this dependence
for each event helicity combination,
$\tau^-_{\rm L}\tau^+_{\rm R}$ and
$\tau^-_{\rm R}\tau^+_{\rm L}$.
The likelihood function for each event
includes contributions from both event helicities
according to the measured $\tau$ polarization~\cite{polar}
as a function of the $\tau^-$ polar angle and $\sqrt{s}$.

The fitted lifetime is sensitive to the $d$ resolution
assumed in the likelihood function.
The $d$ resolution function for each daughter track
is taken to be the sum of three
Gaussian functions whose widths and amplitudes are
parametrized~\cite{alephiii}
in terms of momentum, polar angle, and
the $d$ uncertainty calculated by the track fitting program.
The parameters of the resolution core and near tails
are obtained from studies of
$e^+e^-\rightarrow e^+e^-$,
$e^+e^-\rightarrow\mu^+\mu^-$,
$\gamma\gamma\rightarrow e^+e^-$,
and $\gamma\gamma\rightarrow\mu^+\mu^-$
events in the data.
The parameters of the far tails,
dominated by
photon conversions (which can cause pattern recognition errors)
and nuclear interactions
in hadronic $\tau$ decays,
are extracted from a sample of simulated $\tau^+\tau^-$ events.
The rms of the core is typically 25 to $100\,\mu{\mathrm{m}}$.
The second (third) Gaussian function contains
roughly 5\% ($0.2\%$) of the tracks
and is 3.3 (${\sim}20$) times wider than the core.
Finally, the small smearing in $\delta$
related to the size of the luminous region
(typically $5\,\mu{\mathrm{m}}$ rms)
is taken into account in the likelihood function.

\begin{figure}[p]
\vbox to \textheight{\vfill%
\epsfysize=95mm
\begin{center}\mbox{\epsffile{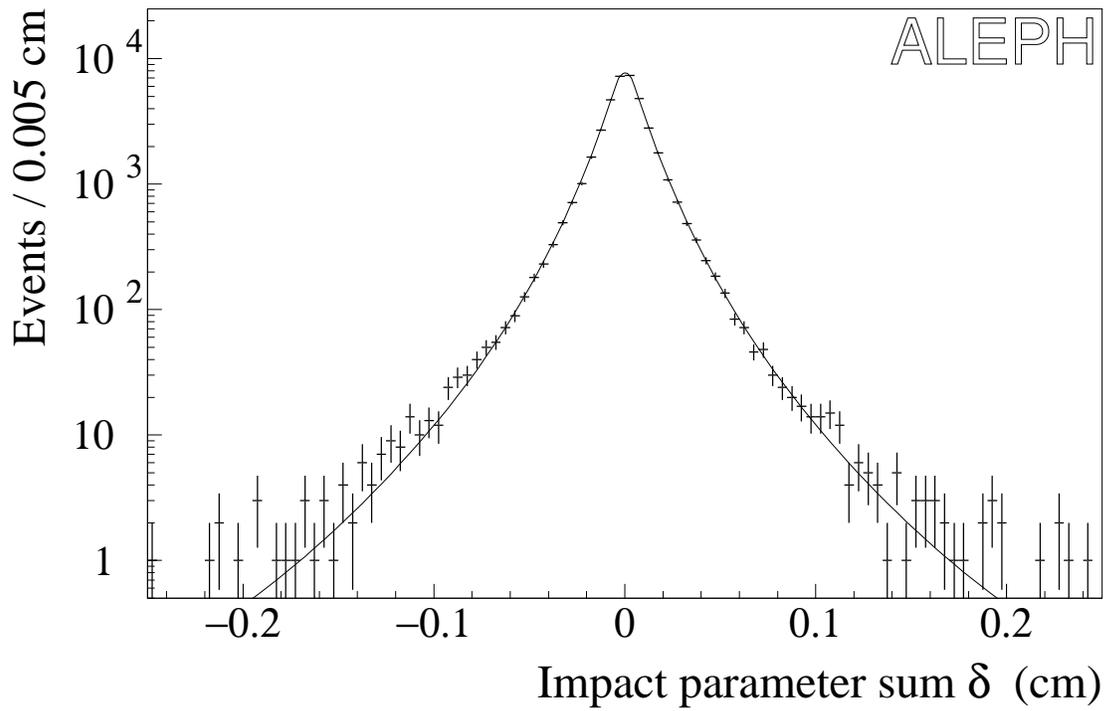}}\end{center}
\caption[]{\label{f:mips}%
Impact parameter sum distribution for 1993 and 1994 data.
The curve represents the sum of the functions obtained
from separate fits to the two data samples.}
\vfill}%
\end{figure}

The event selection~\cite{alephiii} yields 40271 candidate events.
A few remaining events with poorly measured tracks
are removed by means of a cut on the event confidence level (CL).
The CL of an event is defined to be the
integrated probability density for the event to have
a reconstructed $|{\delta}|$
equal to or larger than the observed value;
a mean $\tau$ lifetime of $293.7\,{\mathrm{fs}}$~\cite{alephiii}
is assumed in this calculation.
The requirement $\mathrm{CL} > 0.01\%$ removes
29 events from the sample.
The fits to the data yield an (uncorrected) mean $\tau$ lifetime
of $289.1 \pm 1.8 \,{\mathrm{fs}}$.
Figure~\ref{f:mips} shows the $\delta$ distribution
for the data.

The same procedure,
including the parametrization of the $d$ resolution,
is followed with Monte Carlo
$e^+e^-\rightarrow\ell^+\ell^-$ and 
$\gamma\gamma\rightarrow\ell^+\ell^-$ events
($\ell = e$, $\mu$, $\tau$)
generated at $\sqrt{s} = 91.25\,{\mathrm{GeV}}$.
The different dimensions of the luminous region in
1993 and 1994 are taken into account in the fits
to the Monte Carlo events.
The possible lifetime bias in the method is determined
from a comparison of the fitted lifetime with
the generated value;
the resulting bias is $(+0.24 \pm 0.63)\%$,
where the uncertainty is from Monte Carlo statistics.
The systematic error associated with the use of
peak Monte Carlo events to simulate off-peak data is negligible.

The calculated biases are used to correct
the results of the fits to the data;
the Monte Carlo fit uncertainty is therefore treated as
a systematic uncertainty on the measured lifetime.
The systematic biases and uncertainties are
summarized in Table~\ref{t:mips}.
\begin{table}[t]
\caption[]{\label{t:mips}%
Systematic biases and uncertainties in the MIPS analysis.}
\begin{center}
\begin{tabular}{@{\extracolsep{8mm}}lr}%
\hline\noalign{\smallskip}%
Source & Bias and uncertainty (\%) \\
\noalign{\smallskip}\hline\noalign{\smallskip}
$\tau^+\tau^-$ Monte Carlo bias            &$ +0.24 \pm 0.63$ \\
$d$ resolution parametrization             &$    {} \pm 0.95$ \\
Variation of CL cut                        &$    {} \pm 0.25$ \\
Transverse $\tau$ polarization correlation &$ +0.22 \pm 0.22$ \\
Branching fractions                        &$ -0.06 \pm 0.13$ \\
Backgrounds                                &$ -0.36 \pm 0.09$ \\
\noalign{\smallskip}\hline\noalign{\smallskip}               
Total                                      &$ +0.04 \pm 1.20$ \\
\hline\end{tabular}
\end{center}
\end{table}

The systematic uncertainty associated with the
$d$ resolution parametrization
includes contributions from
(1) the limitations of the resolution model,
estimated from a comparison of the parameters derived from
simulated $\tau^+\tau^-$ vs.\ $e^+e^-$ and $\mu^+\mu^-$ events;
(2) the statistical uncertainty on the parameters
measured from real $e^+e^-$ and $\mu^+\mu^-$ events;
(3) the statistical uncertainty on the parameters of the
far tails, measured from simulated $\tau^+\tau^-$ events; and
(4) the simulation accuracy of the far tail parameters,
estimated from a comparison of real and simulated
$e^+e^-$, $\mu^+\mu^-$, and $\mbox{\sl q}\bar{\mbox{\sl q}}$ events.
An additional test of the resolution
in the $\tau^+\tau^-$ sample is performed
in which the CL cut value is varied between 0 and 0.5\%.
The resulting variations of the fitted lifetime
in data and Monte Carlo agree
within the expected statistical fluctuations;
the assigned systematic uncertainty, $0.25\%$,
reflects the sensitivity of the test.
Any remaining detector alignment errors are implicitly
taken into account in the $d$ resolution parametrization.

The correlation of the transverse polarizations of the
$\tau^+$ and $\tau^-$
is not simulated in the event generator~\cite{koralz}
used for the final lifetime bias determination,
so a special generator~\cite{koralb}
without initial or final state radiation
is used to determine the effective bias due to this correlation,
$(+0.22 \pm 0.22)\%$.
The uncertainty associated with the longitudinal $\tau$ polarization
is negligible.

Each $\tau$ decay mode produces 
a different impact parameter distribution.
The differences between
the measured $\tau$ branching fractions~[\hbox{15},\hbox{16}]
and those used in the Monte Carlo simulation
are expected to yield a bias of $-0.06\%$
on the measured lifetime.
The uncertainties on the measured branching fractions
correspond to a lifetime uncertainty of $0.13\%$.

The bias due to background events is predicted
from Monte Carlo simulation to be
$(-0.36 \pm 0.09)\%$,
where the uncertainty includes a systematic contribution of $25\%$,
estimated from a comparison of
the real and simulated distributions of the discriminating variables
used in the $\tau^+\tau^-$ event selection.
The reaction $\gamma\gamma\rightarrow\ell^+\ell^-$
is the dominant source of contamination in the
1-1 sample,
amounting to $0.27\%$.
The contamination from cosmic rays is measured to be
less than 0.01\%.

The net systematic bias is $(+0.04 \pm 1.20)\%$.
The $\tau$ lifetime result, corrected for biases, is
\begin{equation}
\tau_\tau = 289.0 \pm 1.8 \,{\mathrm{(stat)}} 
              \pm 3.5 \,{\mathrm{(syst)}} \,{\mathrm{fs}}.
\end{equation}

\section{Impact parameter difference analysis}
\label{s:ipd}%

The 1-1 topology events are also analyzed
with the IPD method~\cite{alephi}.
In this method the following quantities are determined for each event:
\begin{equation}
\begin{array}{r@{\ }c@{\ }l}
Y &=& d_+ - d_-\; , \\
X &=& \displaystyle\frac{\bar{p}_\tau(\sqrt{s})}{\bar{p}^0_\tau}
\,{\scriptstyle \Delta}\mskip-1.5mu \phi \,\sin\theta\; ,
\end{array}
\end{equation}
where $\bar{p}_\tau(\sqrt{s})$ is the mean $\tau$ momentum,
determined from Monte Carlo simulation after
all event selection criteria are applied,
$\bar{p}^0_\tau = 45.40\,{\mathrm{GeV}}/c$
is the mean $\tau$ momentum at
$\sqrt{s} = 91.25 \,{\mathrm{GeV}}$,
${\scriptstyle \Delta}\mskip-1.5mu \phi = \phi_+ - \phi_- \pm \pi$
is the acoplanarity of the two daughter tracks,
and $\theta$ is taken to be the polar angle of
the event thrust axis,
calculated from the reconstructed charged and neutral particles.
No estimate of the $\tau$ direction is needed to
determine ${\scriptstyle \Delta}\mskip-1.5mu \phi$.
For a given value of $X$,
the expected value of $Y$ is given by~\cite{alephi}
\begin{equation}
\langle Y \rangle \ = \ 
\left[ \frac{\bar{p}^0_\tau}{m_\tau} \tau_\tau \right] X \; ,
\label{eq:ipd}
\end{equation}
i.e., the slope of $\langle Y \rangle\,{\mathrm{vs.}}\,X$
is equal to the mean $\tau$ decay length in the laboratory frame.
This relation holds in the approximation that
the $\tau^+$ and $\tau^-$
are back to back in the $r$-$\phi$ projection
and the $\tau$ decay angles are small in the lab.

Equation~\ref{eq:ipd} is not satisfied for 
radiative ${\mathrm{Z}}\rightarrow\tau^+\tau^-$ events;
such events are rejected by a cut on
the invariant mass of the charged daughter and the photon candidates
in each event hemisphere.
The parameters of the line 
$\langle Y \rangle = a_0 + a_1 X$
are then extracted by means of
an unbinned least-squares fit,
and the mean $\tau$ lifetime is computed
from the fitted slope $a_1$.
Event $i$ is weighted by $1/(\Delta Y_i)^2$ in the fit,
where $\Delta Y_i$,
the expected rms smearing on $Y_i$,
includes contributions from the
estimated tracking resolution for event $i$,
the size of the luminous region (a function of $\phi_{\pm}$),
and the natural spread of $Y_i$ (a function of $X_i$).
The fit range $\left| X \right| < 0.18$ is chosen in order to
reduce the effect of 
mismeasured tracks,
radiative events,
and background from two-photon interactions;
38120 events enter the fit.
An iterative trimming procedure is used
to remove events with fit residuals in $Y$
greater in magnitude than 
$\Delta_{\mathrm trim} = 0.137\,{\mathrm{cm}}$,
most of which contain mismeasured tracks.
This procedure removes 78 events.

\begin{figure}[p]
\vbox to \textheight{\vfill%
\epsfysize=150mm
\begin{center}\mbox{\epsffile{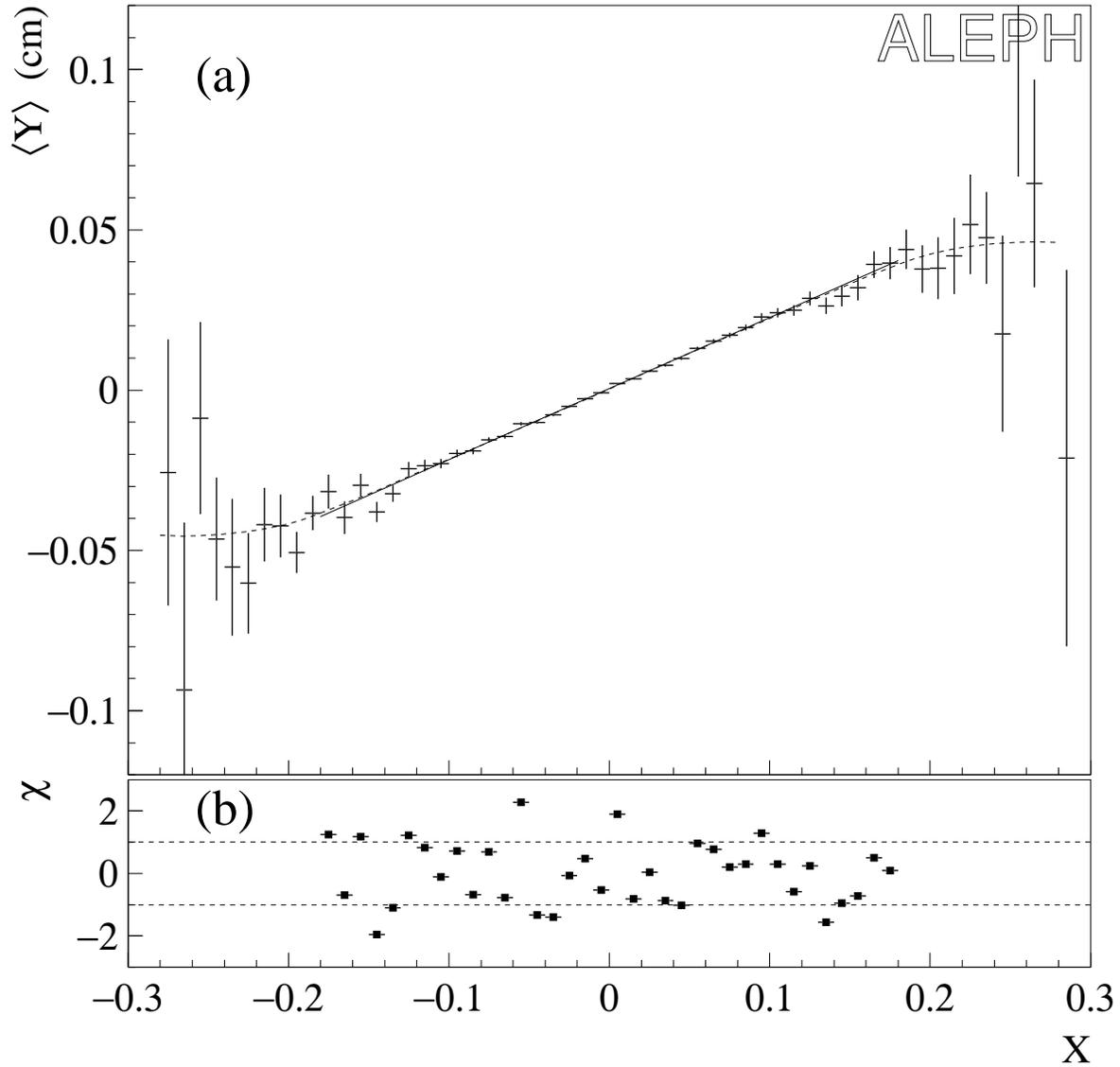}}\end{center}
\caption[]{\label{f:ipd}%
(a) $\langle Y \rangle\,{\mathrm{vs.}}\,X$ for the 1993 and 1994 data.
The solid line represents the fit results.
The dashed curve shows the shape of 
$\langle Y \rangle\,{\mathrm{vs.}}\,X$
for simulated $\tau^+\tau^-$ and background events.
(b) Plot of pulls
(deviation from fitted line divided by uncertainty).}
\vfill}%
\end{figure}

The fits to the data yield
$a_1 = +0.2218 \pm 0.0023 \,{\mathrm{cm}}$ and
$a_0 = +0.0004 \pm 0.0001 \,{\mathrm{cm}}$ (Fig.~\ref{f:ipd});
the $\chi^2$ per degree of freedom is 1.02,
implying that $\Delta Y_i$ is correctly parametrized.
The small positive offset in $a_0$ is caused by
bremsstrahlung and other track measurement errors
and agrees with the value predicted from Monte Carlo events.

Simulated events are used to study and correct for the bias
in the determination of the mean $\tau$ decay length
from the $Y\,{\mathrm{vs.}}\,X$ distribution.
The dependence of $a_1$ on $\tau_\tau$ is influenced by
the following effects:
(1) The selection procedure may introduce a bias on the lifetime.
(2) Surviving radiative events violate the assumption
that the $\tau^+$ and $\tau^-$ are back to back;
they also cause the mean $\tau$ momentum 
to depend slightly on $\left| X \right|$.
(3) There is a bias associated with the approximation that
the $\tau$ decay angles are small in the lab system.
(4) Tracking errors on $d$ and $\phi$ introduce a positive
bias on $a_1$,
which is reduced by the trimming in the fit.
(5) Background events affect the fitted slope.
These biases and all systematic uncertainties
are given in Table~\ref{t:ipd}. 
The net bias is
$(-0.29 \pm 0.20)\%$,
where the uncertainty is from Monte Carlo statistics.
\begin{table}
\caption[]{\label{t:ipd}%
Systematic biases and uncertainties in the IPD analysis.
The first uncertainty is from Monte Carlo statistics
and the second is systematic.}
\begin{center}
\begin{tabular}{@{\extracolsep{8mm}}lr}%
\hline\noalign{\smallskip}
Source & Bias and uncertainty (\%) \\
\noalign{\smallskip}\hline\noalign{\smallskip}
Selection bias                         & \rlap{$-0.02 \pm 0.11$}%
\setbox9=\hbox{$+0.00 \pm 0.00 \pm 0.00$}\hbox to\wd9{\hfil} \\
Radiative events                       & $-0.47 \pm 0.03 \pm 0.09$ \\
Small decay angle approximation        & \rlap{$-0.12 \pm 0.01$}%
\setbox9=\hbox{$+0.00 \pm 0.00 \pm 0.00$}\hbox to\wd9{\hfil} \\
Tracking resolution and trimming       & $+0.54 \pm 0.16 \pm 0.35$ \\
Backgrounds                            & $-0.22 \pm 0.04 \pm 0.06$ \\
Detector alignment                     & $ {}            \pm 0.15$ \\
\noalign{\smallskip}\hline\noalign{\smallskip}
Total                                  & $-0.29 \pm 0.20 \pm 0.40$ \\
\hline
\end{tabular}
\end{center}
\end{table}

In addition to the statistical errors
from the Monte Carlo simulation,
the following systematic errors are considered.
The simulation of final state radiation
in selected ${\mathrm{Z}}\rightarrow\tau^+\tau^-$ decays,
relevant to Bias~2 above,
is verified in an analysis of
isolated photons in data and Monte Carlo.
The lifetime uncertainty associated with
the simulation of the tracking resolution,
including multiple scattering,
bremsstrahlung,
and nuclear interactions,
is evaluated from studies of
the $\tau^+\tau^-$ events
as well as
Bhabha, dimuon, and two-photon events
in data and Monte Carlo.
A systematic uncertainty of $25\%$ is assigned
on the bias from each background source.
The uncertainty associated with detector alignment errors 
is taken to be half the effect of the $d$ offset corrections.
The quadratic sum of these systematic uncertainties
is $0.40\%$.
The beam position and size contribute
a negligible systematic uncertainty.
As a check of the procedure,
the $\Delta_{\mathrm trim}$ value and 
the fit range in $X$ are varied;
the resulting variations in the fitted slope
are consistent with those observed for Monte Carlo events
and no additional systematic uncertainty is assumed.
The total systematic uncertainty,
including the contribution from Monte Carlo statistics,
is $0.45\%$.

The statistical uncertainty on $\tau_\tau$ is multiplied by $1.03$
in order to take into account the small dependence
of the trimming bias on $\tau_\tau$.
The uncertainty on $\bar{p}^0_\tau$ is negligible.
The corrected $a_1$ value corresponds to
\begin{equation}
\tau_\tau =
290.4 \pm 3.2 \,{\mathrm{(stat)}} \pm 1.3
\,{\mathrm{(syst)}} \,{\mathrm{fs}}.
\end{equation}

\section{Decay length analysis}
\label{s:dl}%
The DL method~\cite{alephiii} is used to measure the mean lifetime
of $\tau$'s decaying into three charged tracks.
Three-prong hemispheres with $\Sigma q = \pm1$
are selected from the basic $\tau^+\tau^-$ sample.
The event sphericity axis is calculated
from the reconstructed charged and neutral particles
for each event containing a candidate decay.
The three charged tracks are required to point
within $18^\circ$ of this axis.
Neutral particles outside of this cone are discarded
and the sphericity axis is recalculated.
This procedure avoids the large error on the $\tau$ direction
(and consequent lifetime bias)
which can occur in radiative events.

Decays with an identified electron are rejected
to reduce contamination from photon conversions.
The decay vertex is fitted from the full
three-dimensional track information provided by the detector.
A candidate is retained only if
the vertex fit gives a $\chi^2$ CL greater than $3\%$.
This cut rejects 38\% of the candidates in data
and 31\% in Monte Carlo.
Most of the rejected decays contain a VDET hit that is assigned
to the wrong track.

\begin{figure}[p]
\vbox to \textheight{\vfill%
\epsfysize=95mm
\begin{center}\mbox{\epsffile{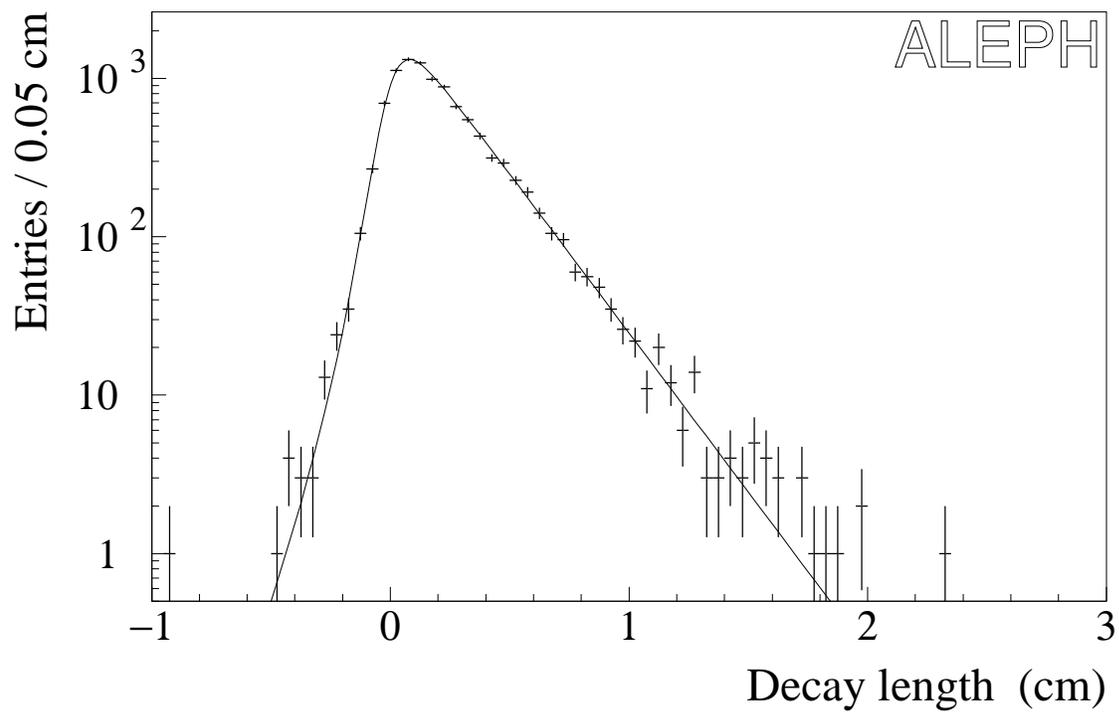}}\end{center}
\caption[]{\label{f:dl}%
Decay length distribution for 1993 and 1994 data.
The curve represents the sum of the functions
obtained from separate fits to the two data samples.}
\vfill}%
\end{figure}

For each candidate decay,
the $\tau$ flight distance is evaluated
by means of a least-squares fit
in which the $\tau$ production and decay points are free to vary.
The position and size of the luminous region and
the position and uncertainty of the fitted decay vertex
enter the fit;
the $\tau$ flight direction is constrained to be parallel,
within an uncertainty of typically $15\,{\mathrm{mrad}}$,
to the event sphericity axis,
calculated as described above.
The $\tau$ direction uncertainty is parametrized
from simulated events
as a function of the event sphericity;
events with low sphericity tend to have low neutrino momenta
and hence a smaller $\tau$ direction uncertainty.
The uncertainty on the fitted decay length
is required to be less than $0.3\,{\mathrm{cm}}$,
whereas the typical decay length resolution is $0.06\,{\mathrm{cm}}$.
Decay candidates with a fitted decay length greater than
$3\,{\mathrm{cm}}$ are discarded.
This requirement removes two candidates
whose reconstructed vertices coincide with the beam pipe
(radius $5.4\,{\mathrm{cm}}$).
Figure~\ref{f:dl} shows the decay length distribution for
the remaining $10076$ candidates.

The mean decay length $\langle\ell\rangle$
is extracted from the decay length distribution
by means of a maximum likelihood fit.
The probability function is taken to be the convolution of
a decreasing exponential with a Gaussian resolution function.
The slope of the exponential is adjusted for each event
according to the LEP energy,
such that the fitted $\langle\ell\rangle$
corresponds to $\sqrt{s} = 91.25 \,{\mathrm{GeV}}$.
The decay length uncertainties,
calculated event by event,
are multiplied by a global scaling factor $k$
which is free to vary in the fit.
The fits to the data yield
$\langle\ell\rangle = 0.2174 \pm 0.0023 \,{\mathrm{cm}}$ and
$k = 1.20 \pm 0.02$.

The bias on the mean decay length,
calculated from $\tau^+\tau^-$ Monte Carlo events,
is $(-0.81 \pm 0.38)\%$,
where the uncertainty is from Monte Carlo statistics.

A Monte Carlo study shows that
background events from 
${\mathrm{Z}}\rightarrow\mbox{\sl q}\bar{\mbox{\sl q}}$
have a mean decay length consistent with zero
and yield a bias of $(-0.26 \pm 0.09)\%$;
this uncertainty reflects the Monte Carlo statistics
and a $25\%$ systematic contribution.
The contamination from one-prong $\tau$ decays
with converted photons is less than $0.01\%$.

The effects of pattern recognition errors are studied by
varying the vertex $\chi^2$ CL cut between 1\% and 5\%;
a systematic uncertainty of $0.42\%$ is assigned,
based on the observed variations in the mean decay length
in data and Monte Carlo.
If the CL cut is placed below 1\% a large increase in $k$
is observed in data and Monte Carlo,
indicating the presence of non-Gaussian tails
in the decay length resolution.

The use of a double-Gaussian parametrization
of the decay length resolution
yields a negligible change (${<}0.05\%$)
in the fitted mean decay length.
The systematic uncertainty associated with the
parametrization of the $\tau$ direction uncertainty
is negligible.
The uncertainties on the detector alignment parameters
correspond to a negligible error on $\langle\ell\rangle$.
The beam position and size contribute
a negligible systematic uncertainty.

The systematic biases and uncertainties are
listed in Table~\ref{t:dl}.
The total bias is calculated to be $(-1.07 \pm 0.57)\%$.
A correction for this bias is applied to
the fitted mean decay length.
The mean momentum of selected $\tau$'s in Monte Carlo events
at $\sqrt{s} = 91.25\,{\mathrm{GeV}}$ is $45.24\,{\mathrm{GeV}}/c$;
this value is used to convert the mean decay length
to a mean proper lifetime:
\begin{equation}
\tau_\tau =
287.9 \pm 3.1 \,{\mathrm{(stat)}} \pm 1.6
\,{\mathrm{(syst)}} \,{\mathrm{fs}}.
\end{equation}
\begin{table}[t]
\caption[]{\label{t:dl}%
Systematic biases and uncertainties in the DL analysis.}
\begin{center}
\begin{tabular}{@{\extracolsep{8mm}}lr}%
\hline
\noalign{\smallskip}
Source & Bias and uncertainty (\%) \\
\noalign{\smallskip}\hline\noalign{\smallskip}
$\tau^+\tau^-$ Monte Carlo bias         & $ -0.81 \pm 0.38$ \\
Backgrounds                             & $ -0.26 \pm 0.09$ \\
Pattern recognition errors              & $    {} \pm 0.42$ \\
\noalign{\smallskip}\hline\noalign{\smallskip}                   
Total bias and uncertainty              & $ -1.07 \pm 0.57$ \\
\noalign{\smallskip}
\hline
\end{tabular}
\end{center}
\end{table}

\section{Conclusions}
\label{s:concl}%
The procedure of~\cite{combine} is used to
determine the optimum weights for
averaging the measured lifetimes from the
MIPS, IPD, and DL analyses.
Correlations among the statistical and systematic errors are
taken into account.
The correlation coefficient of the MIPS and IPD statistical errors
is calculated to be $0.44 \pm 0.04$ in Monte Carlo events.
The combined result for the 1993 and 1994 data is
\begin{equation}
\tau_\tau =
289.0 \pm 1.9 \,{\mathrm{(stat)}} \pm 1.3
\,{\mathrm{(syst)}} \,{\mathrm{fs}},
\end{equation}
with $\chi^2 = 0.29$ for 2 degrees of freedom
(${\mathrm{CL}} = 87\%$).

The three new results are combined with
the previously published ALEPH measure\-ments,
including those obtained with the MIPS, IPD, DL, and other methods
from data samples collected in 1989--1992~[\hbox{1}--\hbox{3}]
and with the 3DIP method
from the 1992--1994 data~\cite{alephthreedip}.
The statistical correlations involving 3DIP are
detailed in~\cite{alephthreedip}.
The combined ALEPH result is
\begin{equation}
\tau_\tau =
290.1 \pm 1.5 \,{\mathrm{(stat)}} \pm 1.1
\,{\mathrm{(syst)}} \,{\mathrm{fs}},
\end{equation}
with $\chi^2 = 9.1$ for 15 degrees of freedom
(${\mathrm{CL}} = 87\%$).
This result,
the most precise measurement of the mean $\tau$ lifetime,
is consistent with other recent measurements~\cite{recent}.

The ALEPH measurements of
the $\tau$ lifetime and branching fractions
may be used 
to test lepton universality.
For $B(\tau\rightarrow e\nu\bar{\nu}) =
(17.79 \pm 0.12 \pm 0.06) \%$~\cite{alephlbf},
$B(\tau\rightarrow\mu\nu\bar{\nu}) =
(17.31 \pm 0.11 \pm 0.05) \%$~\cite{alephlbf},
and other quantities from~\cite{rpp},
the ratios of the effective coupling constants~\cite{marciano} are
\begin{eqnarray}
\frac{g_\tau}{g_\mu} &=& 1.0004 \pm 0.0032 \pm 0.0038 \pm 0.0005 
\label{eq:gtgm} \\
\noalign{\noindent and}
\frac{g_\tau}{g_e} &=&   1.0007 \pm 0.0032 \pm 0.0035 \pm 0.0005,
\label{eq:gtge}
\end{eqnarray}
where the first uncertainty is from the $\tau$ lifetime,
the second is from the $\tau$ leptonic branching fraction
($B(\tau\rightarrow e\nu\bar{\nu})$ in Eq.~\ref{eq:gtgm} and
$B(\tau\rightarrow\mu\nu\bar{\nu})$ in Eq.~\ref{eq:gtge}),
and the third is from the $\tau$ mass.
The measured ratios are consistent with the hypothesis of
lepton universality.

\section*{Acknowledgements}
We wish to thank our colleagues in the CERN accelerator divisions
for the successful operation of LEP.
We are indebted to the engineers and technicians in all our
institutions for their contribution to the excellent performance
of ALEPH.
Those of us from non-member countries thank CERN for its hospitality.


%
\end{document}